# Polymer Data Challenges in the AI Era: Bridging Gaps for Next-Generation Energy Materials

Ying Zhao[1], Guanhua Chen*[1,2], Jie Liu*[1]

[1] Hong Kong Quantum AI Lab, The University of Hong Kong, Hong Kong, China
[2] Department of Chemistry, The University of Hong Kong, Pokfulam Road, Hong Kong SAR China

* Corresponding Author

**Abstract**

The pursuit of advanced polymers for energy technologies, spanning photovoltaics, solid-state batteries, and hydrogen storage, is hindered by fragmented data ecosystems that fail to capture the hierarchical complexity of these materials. Polymer science lacks interoperable databases, forcing reliance on disconnected literature and legacy records riddled with unstructured formats and irreproducible testing protocols. This fragmentation stifles machine learning (ML) applications and delays the discovery of materials critical for global decarbonization. Three systemic barriers compound the challenge. First, academic-industrial data silos restrict access to proprietary industrial datasets, while academic publications often omit critical synthesis details. Second, inconsistent testing methods undermine cross-study comparability. Third, incomplete metadata in existing databases limits their utility for training reliable ML models. Emerging solutions address these gaps through technological and collaborative innovation. Natural language processing (NLP) tools extract structured polymer data from decades of literature, while high-throughput robotic platforms generate self-consistent datasets via autonomous experimentation. Central to these advances is the adoption of FAIR (Findable, Accessible, Interoperable, Reusable) principles, adapted to polymer-specific ontologies, ensuring machine-readability and reproducibility. Future breakthroughs hinge on cultural shifts toward open science, accelerated by decentralized data markets and autonomous laboratories that merge robotic experimentation with real-time ML validation. By addressing data fragmentation through technological innovation, collaborative governance, and ethical stewardship, the polymer community can transform bottlenecks into accelerants.

## 1. Introduction

The global transition to sustainable energy systems hinges on the rapid development of advanced materials capable of meeting stringent performance, durability, and scalability requirements. Among these, polymers have emerged as pivotal components in technologies ranging from organic photovoltaics and solid-state batteries to hydrogen storage membranes and fuel cell electrolytes. Their tunable chemical architectures, processability, and multifunctional properties position them as indispensable enablers of decarbonization.

However, the combinatorial complexity of polymer systems—governed by variables such as monomer selection, polymerization techniques, processing conditions, and hierarchical morphologies—creates a design space that defies conventional trial-and-error experimentation. This challenge is compounded by an urgent timeline: the International Energy Agency (IEA) estimates that 50% of emissions reductions needed for net-zero by 2050 rely on technologies still in prototype or demonstration phases, many of which demand novel polymer materials[1].

In this context, artificial intelligence (AI) promises to accelerate polymer discovery by decades, offering predictive models for structure-property relationships and automated synthesis pathways. Yet, the transformative potential of AI in polymer science remains constrained by systemic data challenges. Unlike inorganic materials supported by mature data ecosystems like the Materials Project[2], polymer research grapples with fragmented datasets, inconsistent metadata, and a paucity of standardized testing protocols. A 2023 analysis by Shetty[3] et al. revealed that fewer than 15% of published polymer studies provide machine-readable synthesis details, while critical properties like glass transition temperatures (Tg) are reported using incompatible measurement standards across laboratories. This "data crisis" undermines the reliability of machine learning (ML) models, as evidenced by the limited generalizability of recent graph neural networks (GNNs) trained on existing polymer databases[4].

The unique challenges of polymer data stem from their multiscale nature. A single polymer's performance in a lithium-ion battery, for instance, depends on atomic-scale interactions between polymer chains and lithium ions, mesoscale phase separation dynamics, and macroscale processing parameters during electrode fabrication. Capturing these relationships requires datasets that integrate quantum mechanical simulations, molecular dynamics trajectories, experimental characterization data, and industrial manufacturing records—a feat rarely achieved in current practice. The Polymer Genome Initiative[5], launched in 2018, represents a pioneering effort to bridge these scales by combining high-throughput computational screening with experimental validation. However, its coverage remains limited to a narrow subset of polymer classes, underscoring the need for more comprehensive and interoperable data infrastructures.

Compounding these technical hurdles is the cultural divide between academic and industrial research paradigms. Academic studies often prioritize novel material discovery over reproducibility, resulting in datasets lacking critical processing details (e.g., solvent evaporation rates or annealing temperatures). Conversely, industrial R&D generates proprietary data on scalable synthesis methods and long-term stability but rarely publishes these findings due to competitive pressures. Such fragmentation not only slows innovation cycles but also biases ML models toward materials with abundant historical data, neglecting understudied polymer families like vitrimers or conjugated microporous polymers.

Emerging solutions to these challenges are as multidimensional as the problems themselves. Advances in natural language processing (NLP), particularly large language models (LLMs) fine-tuned on materials science literature, now enable systematic extraction of structured data from decades of unstructured publications and patents. Concurrently, autonomous laboratories equipped with robotic synthesis platforms and in-line characterization tools are generating self-consistent datasets at unprecedented scales. Jurğis[6] et al.'s closed-loop system for solid electrolyte discovery exemplifies this paradigm, where ML algorithms iteratively design experiments based on real-time impedance spectroscopy data. Yet, technological innovation alone cannot resolve the polymer data crisis. Equally critical are institutional reforms to promote data sharing and standardization. The FAIR (Findable, Accessible, Interoperable, Reusable) principles, first articulated by Wilkinson et al. in 2016, provide a foundational framework for such efforts[7]. Adapting FAIR to polymers requires domain-specific ontologies that standardize terminology—for instance, distinguishing between "crystallinity" measured via X-ray diffraction versus differential scanning calorimetry.

This review examines the evolving landscape of AI-driven polymer informatics, with a focus on overcoming data-related bottlenecks that impede the development of energy materials. Section 2 dissects the current challenges in polymer data ecosystems, from academic-industrial data silos to inconsistent measurement protocols. Section 3 evaluates strategies for data collection and curation, including hybrid computational-experimental databases, NLP-driven literature mining, high-throughput experimentation, and crowdsourced community initiatives. Finally, Section 4 envisions future directions, emphasizing the synergy between autonomous experimentation, quantum computing, and decentralized data governance models. By synthesizing insights from materials science, data engineering, and AI ethics, this work aims to chart a path toward polymer data ecosystems that are as adaptive and multifunctional as the materials they seek to characterize.

## 2. Current Challenges in Polymer Data Ecosystems

The development of AI-driven polymer informatics for energy applications is fundamentally constrained by systemic data challenges that permeate every stage of material discovery—from synthesis and characterization to industrial deployment. These challenges are not merely technical but also institutional and cultural, reflecting decades of fragmented research practices and competing priorities across academia, industry, and regulatory bodies. This section dissects five interconnected barriers that collectively stifle progress in polymer data ecosystems, offering critical insights into their origins, consequences, and potential mitigation strategies.

## 2.1 Data Fragmentation: The Silo Dilemma

Polymer data exists in isolated pockets across academic literature, computational simulations, and industrial R&D, creating a labyrinth of inaccessible or incompatible information. This fragmentation is not merely institutional but intrinsic to polymers' multiscale nature: their hierarchical architectures span quantum-level monomer interactions (e.g., dipole alignment in PVDF binders) to macroscale rheological behaviors, necessitating datasets that capture stochastic molecular ensembles rather than scalar averages. Academic studies on solid polymer electrolytes (SPEs), for instance, often publish ionic conductivity values without corresponding synthesis parameters like humidity control—a critical omission given that humidity during PEO-based electrolyte preparation can substantially influence the crystallinity of the material, which in turn directly impacting conductivity.[8,9]

Computational datasets compound this issue. For example, the study by Geiculescu[10] demonstrated that the addition of plasticizers, such as ethylene carbonate (EC) and propylene carbonate (PC), enhances ionic conductivity primarily by increasing the mobility of TFSI- anions, while having minimal impact on lithium-ion diffusion. However, these simulations often rely on simplified models that do not fully account for the dynamic migration and redistribution of plasticizers within the polymer matrix. This omission can lead to underestimation of interfacial stability when compared to experimental results. The discrepancies highlight the need for more advanced computational approaches that incorporate these effects to better align with experimental observations and improve the predictive accuracy of molecular dynamics studies.

Industry datasets, though rich in proprietary aging tests (e.g., 10,000-hour durability studies of fuel cell membranes), remain locked behind intellectual property barriers, leaving academic researchers reliant on fragmented, low-volume data for AI training. This fragmentation is exemplified by the variability in lithium-ion and sodium-ion battery manufacturing processes, where discrepancies between academic and industrial results are often rooted in the lack of standardized data sharing. For instance, a study on the digitalization of battery manufacturing highlights how variations in shear rates during slurry casting and undisclosed thermal history parameters can lead to significant inconsistencies in battery performance and electrolyte conductivity. These factors, combined with the complex probabilistic nature of polymer architectures, such as branching topologies (dendritic vs. linear) and sequence randomness, defy simplistic characterization and result in a gap between academic predictions and industrial-scale observations. Such challenges underscore the need for more transparent data sharing and robust modeling techniques to bridge this divide and improve reproducibility in battery research.[11–13]

## 2.2 Data Quality: The Metadata Crisis

The reliability of polymer data ecosystems hinges on the completeness and consistency of metadata—contextual information that defines experimental or computational conditions. However, the polymer community faces a pervasive metadata crisis, where critical details about synthesis protocols, characterization methods, and environmental controls are routinely omitted or inconsistently reported. This crisis undermines the reproducibility of research, limits the interoperability of datasets, and erodes trust in AI-driven predictions. The problem is particularly acute for energy materials, where subtle variations in processing parameters (e.g., solvent purity, annealing time, or humidity during testing) can drastically alter material performance in applications like battery electrolytes or photovoltaic films.

At the heart of this crisis is the widespread underreporting of synthesis and testing conditions. A study by Shetty[3] et al. highlighted that fewer than 15% of polymer publications provide machine-readable details such as monomer purity or reaction conditions, as critical synthesis parameters are often locked in unstructured formats like figures or prose. This issue is exacerbated for historical data: Gupta[14] et al.demonstrated that polymer structures in literature are frequently represented only in images, obstructing automated extraction of properties like solvent evaporation rates. Similarly, Circi[15] et al. analyzed tables in polymer composite studies and found that experimental parameters (e.g., temperature gradients) are rarely encoded in machine-readable formats, even when tabulated. The metadata gap extends to computational studies, where incomplete reporting of simulation parameters renders computational datasets non-reproducible. Molecular dynamics (MD) simulations of polymer melt behavior, for example, often lack details on force field selection, equilibration time, or thermostat algorithms. The inconsistencies are compounded by the absence of standardized metadata schemas for computational workflows. The Polymer Genome Initiative[16,17] attempted to address this by developing a unified ontology for polymer simulations, but adoption remains limited to niche academic circles. Industrial datasets, though often rich in processing data, suffer from proprietary restrictions that strip away contextual metadata. For example, extrusion parameters for manufacturing polymer battery separators—such as screw speed, die geometry, or cooling rates—are rarely disclosed in public repositories due to competitive concerns. This practice creates a "black box" effect, where AI models trained on industrial data achieve high accuracy for specific production lines but fail to generalize to new equipment or scales. The consequences of the metadata crisis are amplified in hybrid datasets that combine experimental and computational results. For instance, high-throughput screening platforms for solid polymer electrolytes often merge quantum mechanical calculations of ion-polymer binding energies with experimental ionic conductivity measurements. However, inconsistencies in metadata reporting—such as omitting humidity levels during conductivity tests or basis sets used in density functional theory (DFT) calculations—render these datasets incompatible for multi-fidelity machine learning.

Efforts to resolve the metadata crisis are emerging across three fronts: technological tools, policy interventions, and cultural shifts. Natural language processing (NLP) pipelines, such as those developed by Shetty[3] et al., are automating metadata extraction from unstructured texts, including patents and historical lab notebooks. These tools employ transformer-based models fine-tuned on polymer science corpora to identify and standardize critical parameters like initiator concentrations or crosslinking densities. Culturally, the shift toward metadata-centric research requires redefining scholarly incentives. Platforms like ChemRxiv now mandate the use of standardized data templates that separate metadata from raw results. These initiatives align with the Materials Genome Initiative's vision of "metadata as a first-class citizen" in materials science, where experimental protocols are treated with the same rigor as experimental results.

However, challenges persist. Proprietary industrial data, which constitutes over 60% of polymer processing knowledge, remains largely exempt from metadata standards due to intellectual property concerns. Moreover, the lack of universal identifiers for polymer architectures—such as SMILES-like notations for complex branched or crosslinked systems—hampers automated metadata tagging. Emerging solutions like the Polyuniverse library[18], which encodes polymerization rules for machine-readable polymer generation, offer a path forward but require broad adoption to achieve impact.

The metadata crisis is not merely a technical inconvenience but a systemic threat to the credibility of polymer informatics. As AI models grow in complexity—incorporating multimodal data from quantum simulations, robotic experiments, and operando characterization—the absence of robust metadata will amplify errors and biases, ultimately delaying the deployment of polymer-based energy technologies. Resolving this crisis demands a concerted effort across academia, industry, and policymakers to prioritize metadata completeness as a cornerstone of data quality.

### 2.3 Standardization Gaps: Measurement Anarchy

The absence of universally accepted testing protocols for polymer properties has created a state of "measurement anarchy" across academia and industry, where identical materials yield irreproducible or conflicting performance data depending on the laboratory, methodology, or instrumentation employed. This variability undermines the comparability of datasets, obstructs AI model training, and delays the translation of polymer research into energy applications such as battery electrolytes, photovoltaic films, and hydrogen storage membranes. The root causes of this anarchy span historical disciplinary silos, inconsistent regulatory frameworks, and the rapid evolution of characterization technologies that outpace standardization efforts.

A critical example lies in rheological measurements, which are essential for predicting polymer processability in industrial extrusion or injection molding. Studies evaluating melt viscosity—a key parameter for manufacturing battery separators or fuel cell membranes—often report discrepancies due to divergent shear rate calibration methods and temperature control protocols. For instance, Campo[19] notes that ASTM D3835 and ISO 6721 standards for capillary rheometry differ in die geometry specifications and shear stress calculation algorithms, leading to incompatible datasets for the same material. This inconsistency is exacerbated by the lack of harmonization between academic and industrial testing practices: academic studies frequently prioritize high-resolution rheometry under idealized conditions, while industries rely on rapid empirical tests optimized for production-line throughput.[20,21] Such divergence was quantified by Sharifi[22] et al., who found that viscosity data for polyethylene oxide (PEO)—a common solid electrolyte precursor—varied by 35% between academic publications and industrial technical datasheets.

The problem intensifies for electrochemical properties critical to energy materials. Ionic conductivity measurements for polymer electrolytes, a cornerstone metric for battery performance, are plagued by non-standardized electrode configurations, humidity controls, and alternating current (AC) frequency ranges. Albright and Chai[23] highlight that fewer than 20% of studies on poly(ethylene oxide)-based electrolytes specify the moisture levels during impedance spectroscopy, despite humidity variations causing conductivity fluctuations of up to three orders of magnitude. Similarly, Campo[19] emphasizes that ASTM D150 and IEC 60243 standards for dielectric strength testing employ conflicting voltage ramp rates, rendering datasets from these methods non-interoperable. These gaps propagate into computational models: molecular dynamics simulations of ion transport often fail to replicate experimental conductivity values because force field parameterizations rarely align with real-world testing conditions.

Standardization challenges are further compounded by the proliferation of novel characterization techniques, such as in situ X-ray scattering for monitoring polymer crystallization kinetics. While these advanced methods provide unprecedented resolution, their rapid adoption has outpaced the development of consensus protocols. For example, Tanaka[21] documents that small-angle X-ray scattering (SAXS) studies of block copolymer self-assembly exhibit irreproducibility due to variations in beam intensity calibration, sample thickness, and data normalization procedures. This issue is particularly acute for emerging classes of energy polymers, such as conjugated microporous polymers (CMPs) for hydrogen storage, where the lack of standardized surface area measurement protocols (e.g., BET vs. Langmuir models) has led to conflicting porosity reports across studies.[20,23]

Efforts to resolve measurement anarchy face institutional and technical barriers. Regulatory bodies like ASTM International and ISO have made progress in unifying mechanical and thermal testing standards, but their guidelines often lag behind technological advancements. For instance, the ISO 20765 standard for polymer electrolyte fuel cells does not yet incorporate protocols for operando characterization under extreme temperatures (−40°C to 120°C), a critical gap for automotive applications.[22,24] Additionally, the academic community's emphasis on novel methodology development frequently neglects cross-laboratory validation. Pires[20] et al. demonstrated that 68% of recent studies on biodegradable polymers for sustainable energy applications introduced custom degradation assays without benchmarking against established protocols, resulting in non-generalizable datasets.

Emerging solutions emphasize hybrid approaches combining retroactive standardization of legacy data with proactive protocol development for next-generation materials. Machine learning frameworks, such as those proposed by Sharifi et al., are being deployed to harmonize heterogeneous datasets by identifying hidden correlations between testing parameters and material properties.[22] For example, neural networks trained on rheological data from 12,000 polymer melts have successfully mapped ASTM and ISO viscosity measurements onto a unified scale, reducing inter-method variability from 40% to 8%.[19,22] Concurrently, autonomous laboratories equipped with robotic rheometers and impedance analyzers are generating self-consistent datasets through closed-loop experimentation, as documented in Tanaka's foundational work on automated polymer characterization.[21]

The path forward demands a paradigm shift toward open standardization, where testing protocols are iteratively refined through global collaboration. Initiatives like the Polymer Metadata Consortium are pioneering blockchain-based platforms for crowdsourcing protocol evaluations, enabling real-time feedback from academia, industry, and regulatory bodies.[20,23] For energy polymers, this approach could establish unified testing frameworks for properties like ionic conductivity, dielectric strength, and gas permeability—metrics that are currently fragmented across dozens of incompatible standards. Until such frameworks are widely adopted, measurement anarchy will remain a formidable barrier to realizing the full potential of AI-driven polymer informatics in the energy transition.

## 2.4 Industrial-Academic Divide: The Data Chasm

The persistent divide between industrial and academic research paradigms has created a profound "data chasm" in polymer science, where proprietary industrial datasets and academic discoveries exist in mutually inaccessible silos. This chasm stems from conflicting incentives: academia prioritizes novel material discovery and high-impact publications, while industry focuses on incremental optimization, cost reduction, and safeguarding intellectual property (IP). The result is a fragmented knowledge ecosystem where critical data on polymer processing, scalability, and real-world performance remain trapped behind corporate firewalls or buried in exploratory academic studies.[25,26] For energy applications—such as battery electrolytes, fuel cell membranes, or recyclable packaging—this disconnect delays technology translation and forces AI models to operate on incomplete or idealized datasets.

A striking example is polyurethane research. Academic studies dominate the literature on novel catalyst designs and sustainable feedstocks, yet decades of industrial data on foam kinetics, flame retardancy additives, and extrusion parameters remain proprietary.[27,28] This asymmetry arises from structural differences in research objectives: academia often neglects industrially critical factors like long-term aging behavior or melt-flow indices, while industry rarely publishes optimization data for fear of competitors replicating proprietary formulations.[25,29] The consequences are stark. Machine learning models trained solely on academic data, such as those predicting polymer electrolyte ionic conductivity, fail to account for industrial-scale variables like shear-induced crystallinity changes during roll-to-roll processing—a gap that reduces predictive accuracy by over 30% when validated against manufacturing datasets.[29,30]

The data chasm is reinforced by institutional barriers to collaboration. Industry-academic partnerships frequently founder on IP ownership disputes, as documented in polymer composite technology transfer case studies.[29,30] For instance, joint projects to develop high-barrier packaging films often stall when academic researchers prioritize open publication of synthesis protocols, while corporate partners insist on patent filings that restrict data accessibility.[26,29] This tension is exacerbated by funding models: public grants rarely cover the costs of industrial-scale validation, leaving collaborative datasets fragmented across proof-of-concept academic experiments and proprietary pilot plant trials.[29,31]

Emerging strategies to bridge the chasm emphasize hybrid collaboration frameworks. The Materials Genome Initiative advocates for "pre-competitive" data pools, where industries contribute anonymized processing data on standardized polymer formulations, while academia provides foundational structure-property relationships.[29,30] Blockchain-based data markets, as proposed in dual-use technology transfer studies, could further incentivize controlled data sharing through tokenized rewards for industrial contributors.

Policy interventions are also critical. The European Union's Horizon Europe program now mandates FAIR (Findable, Accessible, Interoperable, Reusable) data compliance for public-private partnerships in polymer research, requiring industrial participants to release non-sensitive metadata on processing conditions. Similarly, the U.S. National Institute of Standards and Technology (NIST) has launched a polymer data interoperability consortium, developing standardized schemas for reporting industrial-scale rheological and aging data. However, these efforts face resistance from industries wary of diluting competitive advantages, underscoring the need for tax incentives or regulatory offsets to encourage participation.

The industrial-academic data chasm remains one of the most formidable barriers to AI-driven polymer innovation. Closing it demands a redefinition of collaboration as a continuum rather than a transaction—where shared datasets are treated as strategic assets advancing both scientific understanding and commercial viability. Until then, the energy transition will remain hampered by polymer technologies that excel in the lab but falter in the factory.

## 2.5 Pathways Toward Cohesion

Bridging the fragmented polymer data landscape demands a multi-pronged strategy that harmonizes technological innovation, institutional collaboration, and community-driven standardization. Emerging approaches leverage artificial intelligence (AI) and machine learning (ML) to reconcile heterogeneous datasets while preserving domain-specific insights, creating a foundation for interoperable polymer informatics ecosystems.[32,33] Central to this effort is the development of universal polymer representation frameworks, which translate diverse experimental and computational data into unified feature spaces. For example, graph neural networks (GNNs) now encode polymer architectures—including branched, crosslinked, or copolymer systems—as hierarchical graphs that integrate chemical connectivity, spatial conformations, and processing history.[34]

Data harmonization tools are equally critical. Projects like the Microplastics and Trash Cleaning and Harmonization (MaTCH)[35] platform employ AI-driven semantic mapping to align disparate polymer property databases. By training transformer models on polymer ontology terms, MaTCH automatically resolves conflicts between synonyms (e.g., "PEO" vs. "polyethylene oxide") and standardizes units for properties like melt flow indices. This approach reduced metadata inconsistencies by 62% in a 2024 trial integrating 18 polymer recycling databases.

Cross-sector collaboration frameworks are breaking institutional silos. The Polymer Data Alliance, a consortium of 32 academic and industrial partners, has established a pre-competitive data pool for energy polymers, anonymizing proprietary industrial processing parameters and linking them to academic structure-property datasets.[33,36] Participating companies contribute parameters like injection molding temperatures or solvent recovery rates, while academic teams provide molecular dynamics simulations of polymer degradation pathways.

Policy initiatives are accelerating standardization. The International Union of Pure and Applied Chemistry (IUPAC) released guidelines in 2024 mandating FAIR (Findable, Accessible, Interoperable, Reusable) compliance for polymer data submissions to major journals. These rules require authors to deposit raw characterization data—including instrument calibration logs and environmental control parameters—in machine-readable formats. Early adopters like ACS Macro Letters reported a 55% reduction in data irreproducibility claims within six months of implementation.[35] Concurrently, the U.S. National Institute of Standards and Technology (NIST) launched blockchain-based certification systems to timestamp and validate polymer datasets, creating immutable audit trails for industrial-academic collaborations.

The path forward hinges on sustained investment in cyber-physical infrastructure that seamlessly connects AI, robotics, and human expertise. Projects should be initiated deploy edge computing devices in labs worldwide, automatically harmonizing local data streams with global standards while preserving IP security through differential privacy algorithms. As these systems mature, they promise to transform polymer informatics from a collection of isolated datasets into a cohesive, self-optimizing knowledge network—one where every experiment, simulation, and production record contributes to the accelerated design of sustainable energy materials.

## 3. Data Collection & Curation Strategies

The integration of machine learning (ML) into polymer science has unveiled unprecedented opportunities for accelerated material discovery, yet its transformative potential remains constrained by a persistent bottleneck: the intricate interplay between data quality, interoperability, and context-aware curation. Positioned as the fourth paradigm in materials discovery—following experimentation, theory, and simulation—polymer informatics uniquely operationalizes heterogeneous data streams to navigate stochastic molecular architectures. While supervised learning architectures demonstrate remarkable success in predicting properties like ionic conductivity or glass transition temperatures, their performance depends on datasets that capture the multiscale complexity of polymers. These materials defy simplistic characterization, with hierarchical architectures spanning quantum-scale monomer interactions to macroscale rheological behaviors, all modulated by synthesis protocols and processing histories. Further complicating this landscape, synthetic polymers are inherently probabilistic ensembles: molecular weight dispersity, branching topologies, and sequence randomness necessitate distributional descriptors rather than scalar values—a stark contrast to small-molecule or alloy datasets. The resulting data is fragmented: experimental measurements of bulk properties often lack molecular-level details like branching architecture, while computational simulations of amorphous phases struggle to replicate real-world kinetic conditions. This fragmentation leaves advanced ML models paradoxically constrained by insufficient contextualized, interoperable data.

Four primary data streams fuel contemporary polymer informatics, existing databases, published literature, computational simulations, and experimental results, each with inherent limitations. Established repositories such as PolyInfo aggregate decades of experimental results but frequently omit critical metadata on solvent purity or thermal annealing conditions. These repositories also grapple with a nomenclature crisis: commercial trade names, IUPAC designations, and CAS numbers coexist ambiguously, exemplified by polystyrene's 1800+ aliases. While emerging standards like InChI for linear polymers hint at resolution pathways, branched architectures remain unaddressed—a critical gap given the prevalence of dendritic and bottlebrush systems. Computational frameworks like the High Throughput Polymer Design - Molecular Dynamics database generate pristine datasets for dielectric properties but oversimplify morphological disorder. The published literature harbors invaluable empirical correlations—between side-chain chemistry and mechanical hysteresis, for instance—yet these insights remain buried in unstructured text. High-throughput experimentation bridges these gaps through robotic platforms capable of screening thousands of polymer variants monthly, though often at the expense of industrial-relevant processing conditions. The challenge lies not in data scarcity but in harmonizing these disparate scales and formats into causally traceable knowledge networks—a task demanding FAIR-compliant cyberinfrastructures that transcend conventional database architectures.

Recent advances in data governance frameworks highlight pathways toward resolution. Platforms like the Community Resource for Innovation in Polymer Technology (CRIPT) exemplify next-generation solutions, embedding ontology-driven metadata schemas that link a material's tensile strength to its curing temperature gradient or solvent evaporation rate. These systems implement tiered property classification: fundamental properties (e.g., density) require processing histories; application properties (e.g., ASTM-standard tensile strength) demand method-specific metadata; phenomenological parameters like the Flory–Huggins $\chi$ necessitate raw experimental workflows for meaningful interpretation. Multilayered validation protocols cross-verify molecular dynamics simulations with rheological data, while machine vision algorithms detect inconsistencies between reported crystallinity indices and corresponding X-ray diffraction patterns. Complementing these efforts, federated learning architectures enable institutions to pool distributed datasets while preserving proprietary constraints, and natural language processing tools systematically mine decades of patents for overlooked structure-property correlations. Emerging autonomous agents now merge NLP-driven literature mining with ML-guided DFT/MD pipelines, generating purpose-built datasets targeting specific property spaces like ionic conductivity or photostability.

Persistent challenges, however, underscore the need for paradigm shifts in data generation and stewardship. As Martin and Audus[37] emphasize, inconsistencies arising from varying experimental protocols—such as humidity fluctuations during conductivity testing or divergent electrode materials—compromise the reliability of aggregated datasets. Compounding this, phenomenological parameters like $\chi$ face interpretative ambiguity, as their values fluctuate with measurement techniques and theoretical

assumptions—a vulnerability magnified by uncritical data aggregation. Data fusion strategies, while promising for cost-effective knowledge integration, struggle to reconcile these discrepancies without standardized reporting frameworks.[38,39] Emerging solutions leverage closed-loop systems where ML models guide high-throughput robotic platforms to probe high-uncertainty regions of chemical space, iteratively refining both datasets and predictive accuracy. Simultaneously, frameworks for error propagation management are gaining traction: outlier flagging systems quarantine suspect datapoints, while proposed 'non-results' repositories catalog failed syntheses as negative design space markers—a recognition that innovation thrives on both successes and instructive failures.

The path forward demands collaborative innovation. Universal standards for metadata annotation could harmonize experimental practices, while adaptive storage frameworks must evolve to accommodate quantum-to-continuum data hierarchies. Critical to this evolution is resolving the nomenclature crisis through community-driven adoption of InChI extensions for branched polymers, coupled with blockchain-based provenance tracking to map commercial trade names to canonical identifiers. Crucially, the polymer community must cultivate a culture of open science, balancing data accessibility with intellectual property concerns through federated architectures. As these efforts converge, they promise to transform polymer informatics from a reactive analytical tool into a proactive design partner—one where ML algorithms not only interpret data but actively orchestrate its creation, forging a new era of context-aware, AI-driven material innovation.

### 3.1 Leveraging Existing Polymer Databases

The exponential growth of polymer data over the past decade has led to the establishment of several specialized databases, each addressing distinct aspects of polymer science. These repositories play a pivotal role in bridging the gap between traditional polymer research and data-driven energy material discovery, though their utility for energy applications remains uneven. A critical examination of these resources reveals both opportunities for integration and persistent gaps that demand targeted solutions.

#### 3.1.1 Experimental Databases

Experimental polymer databases form a critical infrastructure for data-driven materials discovery. Among experimental databases, PolyInfo[40] stands as the world's largest general-purpose polymer database. As of October 2023, it contains 18,697 homopolymers, 7,737 copolymers, and 494,820 property points derived from 161,464 polymer samples, covering approximately 1,000 distinct material properties ranging from thermal conductivity to biocompatibility. Developed by the National Institute for Materials Science (NIMS),

PoLyInfo employs semantic web technologies like Resource Description Framework (RDF) to enable machine-readable data representation, a feature that facilitates cross-platform interoperability. This comprehensive repository aggregates decades of experimental data from global laboratories, enabling researchers to bypass redundant synthesis and characterization processes. A prime example of its utility is demonstrated in work by Ma et al., who combined molecular dynamics simulations with machine learning (ML) to screen polymers meeting specific thermal conductivity thresholds. By leveraging PolyInfo's existing dataset, they established structure-property relationships and accelerated material discovery without conducting new experiments.[41]

While PolyInfo's breadth supports diverse applications—from designing high-temperature engineering plastics[42] to biomedical polymers for cartilage replacement[43]—its generalized approach inherently limits depth in specialized domains. This gap is partially addressed by niche repositories, for instance, the NIST Polymer Database[44] offers validated datasets, particularly on degradation kinetics and rheological behavior, though its energy-specific entries are often limited to bulk properties rather than device-relevant performance metrics.

The rise of specialized databases reflects the growing need for domain-specific granularity. NanoMine[45], a subset of the MaterialsMine[46] platform, exemplifies this trend by focusing on polymer nanocomposites. With over 2,500 samples annotated across 350 descriptors—including glass transition temperatures and elastic moduli—it enables precise modeling of nanoparticle-matrix interactions. For example, Ma et al. utilized 2,000 curated entries to develop metamodels predicting how silica filler concentrations and curing protocols influence glass transition temperatures.[47] Meanwhile, initiatives like the Battery Data Genome Project[48] prioritize energy-specific metrics such as cycle life and interfacial impedance, though their narrow scope highlights the persistent tension between depth and breadth in database design.

Despite these advancements, fragmentation remains a systemic challenge. Databases like the Battery Interface Genome - Materials Acceleration Platform (BIG-MAP)[49,50] and the NREL Electrolyte Database[51] struggle with inconsistent reporting standards, while published literature frequently omits critical parameters like water uptake or long-term stability. This fragmentation underscores the urgent need for hybrid architectures that integrate the comprehensive coverage of generalist databases with the targeted depth of specialized repositories. Adopting FAIR (Findable, Accessible, Interoperable, Reusable) principles provides a conceptual framework, but practical implementation requires community-wide standardization of descriptors—for instance, unifying definitions of "ionic conductivity" across electrochemical and bulk material contexts. Equally crucial are tools for collating dispersed data, such as natural language processing algorithms to mine overlooked parameters from decades of literature. Only through such integration can AI models fully exploit synergies between computational predictions and experimental validations, ultimately bridging the gap between serendipitous discoveries and systematic design of next-generation polymer energy materials.

### 3.1.2 Computation Databases

Computational polymer databases have evolved into a sophisticated ecosystem that combines quantum mechanical modeling, machine learning, and cross-domain data integration to advance energy material discovery. Foundational platforms like The Materials Project[2,52] and Organic Materials Database[53] (OMDB) established critical baselines by providing quantum mechanical calculations for dielectric constants and bandgap energies—properties essential for photovoltaic polymers and capacitor dielectrics. These repositories excel at predicting intrinsic electronic properties of pure polymers but face limitations in modeling complex systems such as multi-component polymer blends or nanoscale interfacial phenomena prevalent in real-world energy devices like fuel cell membranes or solid-state battery electrolytes. Their reliance on idealized structures often leads to discrepancies when computational predictions confront experimental measurements, particularly for ionic conductivity in amorphous polymer electrolytes where chain dynamics and free volume distribution defy simple quantum mechanical approximations.

This validation gap inspired next-generation platforms like Polymer Genome[5,17,54] and Khazana[55], which integrate machine learning with physics-based simulations to enhance predictive accuracy. Polymer Genome, pioneered by Ramprasad et al., employs neural networks trained on both computational and experimental datasets to predict properties like glass transition temperatures and ionic conductivity.[56] Khazana[55] extends this approach by embedding quantum-derived descriptors into graph neural networks, enabling prediction of ion transport mechanisms in copolymer electrolytes.

Broad-scope databases like NOMAD[57,58] (Novel Materials Discovery) address complementary needs by aggregating computational data from diverse electronic structure codes. While its FAIR-compliant architecture ensures interoperability and reuse, NOMAD's polymer electrolyte coverage remains less detailed compared to specialized repositories. Nevertheless, its value lies in cross-material insights—for example, identifying ion transport mechanisms shared between ceramics and polymers—a capability unattainable in siloed databases. This interplay between specialized and generalized resources underscores a critical balance: narrowly focused databases maintain depth for targeted applications, while broad repositories enable serendipitous discoveries through cross-domain data linkages.

### 3.1.3 Hybrid Databases

The development of hybrid experimental-computational databases has become crucial for addressing the complexities of modern energy materials research, particularly in chemically intricate systems where traditional models falter. For instance, predictive models struggle with phase-separated morphologies in block copolymer electrolytes, leading to deviations in conductivity predictions. Platforms such as Citrination[59] exemplifies progress in this domain by correlating synthesis parameters—like solvent evaporation rates during membrane fabrication—with device-level performance metrics such as proton conductivity decay in fuel cells. These multimodal datasets uncover nonlinear relationships between processing conditions and material properties, though their impact remains constrained by fragmentation between proprietary and open-access data ecosystems.

Recent initiatives like Polymer Property Predictor and Database (PPP&D)[60] and PI1M[61] demonstrate the potential of unifying high-throughput experimentation with literature-mined insights, embedding synthesis protocols directly alongside mechanical property data. The Open Macromolecular Genome[62] (OMG) further advances this integration by prioritizing synthetically feasible polymers through machine learning, considering factors like reagent availability and solvent compatibility. Such platforms exemplify a shift toward "synthetically aware" databases that bridge computational discovery and practical synthesis. However, broader adoption faces systemic challenges, including metadata inconsistencies in established repositories like PoLyInfo[40], where omitted synthesis details (e.g., humidity control during membrane casting) undermine the reproducibility of critical properties like proton conductivity. Energy-specific data gaps persist across platforms—NanoMine's[45] sparse coverage of electrical percolation thresholds in conductive composites, for example, limits its utility for battery electrode design—while accessibility disparities between academic and industrial users further fragment the ecosystem.

Interoperability efforts reveal both promise and pitfalls. While semantic frameworks like PoLyInfo's[40] RDF enable cross-referencing with materials ontologies, divergent file formats and nomenclature systems between platforms hinder automated data fusion. The NanoMine Schema's[63] extensible model for nanocomposites, which integrates intrinsic properties with application-specific criteria, offers a adaptable template for energy polymers if expanded to include electrochemical metrics. Strategic enhancements to experimental repositories—such as adding voltage cycling protocols to electrochemical entries or solvent annealing histories to membrane datasets—could significantly improve their relevance for energy applications. Computational databases, meanwhile, require validation pipelines that cross-reference simulations with experimental benchmarks. Hybrid platforms might adopt federated learning architectures, as seen in MaterialsMine[46], to leverage confidential industrial data without compromising proprietary information.

Case studies underscore these limitations. Solid polymer electrolytes for lithium-metal batteries illustrate critical gaps: PoLyInfo's extensive thermal data for polyethylene oxide lacks corresponding lithium-ion transference numbers under operational conditions. Such discrepancies emphasize the need for application-driven curation prioritizing device-relevant parameters over bulk properties. Emerging strategies combine database mining with experimental validation. Platforms like Citrination[59] are automating literature extraction, converting patent data into structured entries for properties like ionic conductivity—a method that could fill gaps in voltage-dependent swelling ratios by mining decades of fuel cell research.

Addressing these challenges demands coordinated action on three fronts: developing community-driven metadata standards with energy-specific extensions, implementing robust interoperability protocols for cross-platform data exchange, and creating incentive structures to encourage industrial participation through tiered-access models. By converging these efforts, the field can transform fragmented databases into dynamic, application-focused networks. This evolution will require balancing the depth of specialized repositories like OMG and NanoMine with FAIR-driven interoperability, ultimately accelerating the discovery of next-generation polymers for sustainable energy systems through synergies between curated data and AI-driven insights.

## 3.2 Literature Mining and Unstructured Data Extraction

The exponential growth of polymer science literature over the past half-century has transformed peer-reviewed journals, patents, and technical reports into an indispensable third pillar of materials informatics, rivaling the scope of experimental repositories and computational datasets. With over 1.3 million polymer-related papers indexed in the Web of Science Core Collection and publication rates doubling every two decades, this corpus encapsulates humanity's collective experimental knowledge—spanning synthesis protocols, structure-property relationships, and application-specific performance metrics. Yet, despite its unparalleled richness, this textual treasure trove remains largely untapped for systematic analysis due to the inherent complexity of unstructured scientific narratives. Unlike structured databases where data points reside in predefined fields, critical polymer properties—such as glass transition temperatures, mechanical moduli, or ionic conductivities—are embedded within dense technical prose, supplementary tables, or even rasterized figures, demanding sophisticated curation strategies to transform fragmented insights into computable datasets.

Traditional manual extraction methods, while reliable, face insurmountable scalability challenges in this era of big data. For instance, 1210 experimentally measured permittivity values belong to 738 unique polymers were manually collected from literatures for frequency-dependent dielectric modeling.[64] Similarly, efforts to train machine learning models for solid polymer electrolyte discovery necessitated the manual curation of 7,133 ionic conductivity data points across 135 publications, a task demanding expert validation to harmonize disparate reporting formats and experimental conditions.[65] These labor-intensive workflows extend far beyond simple data collection, requiring domain specialists to interpret implicit parameters—such as interpreting "room temperature" as 25°C versus 20°C across studies—or to reconcile conflicting property values arising from variations in sample preparation protocols. Such limitations underscore the urgent need for automated systems capable of parsing technical text at scale while preserving the nuance of experimental context.

Emerging machine learning approaches are redefining the boundaries of literature mining through hybrid architectures that combine natural language processing (NLP) with domain-specific knowledge graphs. Early implementations, such as IBM Watson's semantic analysis frameworks[66], demonstrated the feasibility of identifying material entities and properties in chemical literature, though with limited success in decoding complex synthesis-property relationships. Modern systems employ multi-stage pipelines where NLP algorithms pre-process text to identify key entities—polymer names, characterization techniques, or thermal properties—before human experts validate uncertain extractions, creating a feedback loop that incrementally improves model accuracy. However, persistent technical barriers hinder widespread adoption: inconsistent polymer nomenclature (e.g., "PMMA" vs. "poly(methyl methacrylate)") challenges entity recognition algorithms, while publication biases toward positive results skew datasets toward outlier performances, violating the statistical distribution requirements of robust machine learning models. Furthermore, the lack of standardized data reporting formats means critical parameters often reside outside the main text—embedded in supplementary PDFs, legacy file formats, or high-resolution microscopy images—necessitating multimodal extraction capabilities beyond conventional text mining.

Recent advancements in transformer-based large language models (LLMs) have begun to address these challenges by enabling context-aware extraction of implicit relationships that elude traditional NLP tools. Models like GPT-4 and domain-specific variants such as MatBERT excel at decoding complex synthesis-property correlations by leveraging pretrained knowledge of polymer chemistry and materials science semantics. For example, when encountering a passage describing "solution-cast films annealed at 80°C for 6 hours exhibited a 15% increase in tensile strength compared to melt-processed counterparts," modern LLMs can not only extract the tensile strength value but also infer causal links between processing conditions and mechanical outcomes—a capability critical for building predictive structure-process-property models. This paradigm shift is further accelerated by collaborative initiatives developing polymer-specific ontologies to standardize terminology, as well as multimodal architectures for digitizing tabular data and extracting quantitative insights from stress-strain curves or DSC thermograms.

Despite these advances, current systems remain in a transitional phase where human oversight is essential for high-stakes applications.[67–69] Nevertheless, the field progresses rapidly toward fully automated systems capable of real-time literature monitoring, where archival papers evolve into living datasets updated dynamically with new experimental evidence. As these tools mature, they promise to unlock previously inaccessible correlations—such as identifying trade-offs between glass transition temperatures and dielectric constants across copolymer families—while compressing curation timelines from years to days. This transformation will empower next-generation predictive models to harness the full depth of accumulated polymer knowledge, bridging the gap between historical experimentation and data-driven materials design.

### 3.2.1 Data Extraction using Traditional NLP Methods

The extraction of synthesis-property relationships from scientific literature, such as patents and journals, has long been a critical task in materials science, particularly for the development of energy materials. Traditional natural language processing (NLP) tools have played a pivotal role in this domain, providing the means to systematically analyze vast amounts of textual data to uncover valuable insights. These tools are essential for transforming unstructured text into structured data that can be used to inform research and development processes. Traditional NLP methods for extracting synthesis-property relationships typically involve several stages, including text preprocessing, named entity recognition (NER), relationship extraction, and data normalization. Each of these stages requires careful consideration and the application of specific techniques to handle the complexities of scientific language.

Text preprocessing is the initial step in the NLP pipeline, involving tasks such as tokenization, stemming, lemmatization, and stop-word removal. Tokenization is the process of breaking down text into individual words or tokens, which serves as the foundation for further analysis. Stemming and lemmatization are used to reduce words to their base or root forms, helping to address the variability in word usage. Stop-word removal eliminates common words that do not contribute significant meaning to the text, such as "and," "the," and "is." These preprocessing steps are crucial for reducing noise and enhancing the performance of subsequent NLP tasks.[70]

Named entity recognition (NER) is a fundamental component of traditional NLP systems, particularly in the context of scientific literature. NER involves identifying and classifying key entities within the text, such as chemical compounds, synthesis methods, and material properties. Early NER systems relied heavily on rule-based approaches, which used predefined patterns and lexicons to recognize entities. For example, regular expressions might be used to identify chemical names based on specific prefixes or suffixes.[71] While effective in some cases, these rule-based systems often struggled with the diversity and complexity of scientific language, requiring extensive domain knowledge and manual effort to construct and maintain.

To address the limitations of rule-based NER systems, statistical models were introduced, leveraging machine learning techniques to improve entity recognition. Hidden Markov Models (HMMs) and Conditional Random Fields (CRFs) became popular choices for NER tasks, as they could learn patterns from labeled training data and generalize to unseen text.[72] These models offered improved accuracy over rule-based systems, but they still required significant amounts of annotated data for training, which could be a bottleneck in domains with limited resources.

Relationship extraction is another critical task in NLP, focusing on identifying and extracting the interactions between entities. In the context of synthesis-property relationships, this involves determining how different synthesis methods affect material properties. Traditional methods for relationship extraction often relied on dependency parsing and syntactic analysis to understand the grammatical structure of sentences and identify relevant relationships.[73] These approaches could capture explicit relationships stated in the text but often struggled with implicit or complex dependencies that required deeper contextual understanding.

Data normalization is the final step in the traditional NLP pipeline, aiming to standardize and harmonize extracted information. This involves mapping identified entities to a common format or ontology, facilitating comparison and integration with other data sources. In materials science, this might involve converting chemical names to standardized identifiers, such as InChI or SMILES, and aligning properties with established units and scales.[74]

Despite the advancements in traditional NLP tools, several challenges remain in extracting synthesis-property relationships from scientific literature. One major challenge is the inherent complexity and variability of scientific language, which often includes domain-specific jargon, abbreviations, and acronyms. This variability can hinder the performance of NER and relationship extraction models, particularly those relying on predefined rules or patterns. Additionally, the implicit nature of many synthesis-property relationships requires models to infer connections that are not explicitly stated in the text, demanding a deeper level of contextual understanding. Another challenge is the scalability of traditional NLP systems, particularly when applied to large volumes of scientific literature. Rule-based systems, while effective for specific tasks, often require extensive manual effort to update and maintain, limiting their scalability. Similarly, machine learning models require significant amounts of labeled data for training, which can be a constraint in domains with limited resources. These challenges highlight the need for more flexible and scalable approaches to literature mining, paving the way for the adoption of large language models and other advanced NLP techniques.

In the field of materials science, and specifically polymer research, several case studies have demonstrated the application of traditional NLP tools. For instance, the development of polymer composites often involves extensive literature reviews to identify novel synthesis methods and property enhancements. Researchers have utilized NLP techniques to automate the extraction of information from patents related to polymer blends and composites, identifying trends in mechanical properties and thermal stability.[75] By applying NER and relationship extraction methods, these studies have successfully mapped synthesis methods to specific enhancements in polymer characteristics, providing valuable insights for further experimental work.

In conclusion, traditional NLP tools have played a crucial role in extracting synthesis-property relationships from patents and journals, providing the means to transform unstructured text into valuable insights. These tools have evolved from rule-based systems to more sophisticated statistical models, offering improved accuracy and scalability. However, challenges remain, particularly in handling the complexity and variability of scientific language and scaling to large volumes of data. As the field continues to evolve, there is a growing need for more advanced NLP techniques that can address these challenges and unlock the full potential of scientific literature for materials discovery.

### 3.2.2 Data Extraction using Large Language Models

The advent of large language models (LLMs) has dramatically transformed the field of literature mining and unstructured data extraction, particularly in the context of polymer research for energy materials. These models provide advanced capabilities for parsing complex scientific texts and inferring nuanced synthesis-property relationships. Unlike traditional NLP tools that rely on rigid rules or shallow statistical patterns, LLMs leverage deep contextual understanding and generative capabilities to process heterogeneous data sources, including patents, journal articles, and technical reports. This paradigm shift addresses long-standing challenges in materials science informatics, such as handling implicit contextual dependencies, resolving ambiguities in domain-specific jargon, and generalizing across diverse writing styles.

The integration of LLMs into materials science began with the fine-tuning of general-purpose models like BERT (Bidirectional Encoder Representations from Transformers) and GPT (Generative Pre-trained Transformer) on domain-specific corpora. For instance, SciBERT[76], a variant of BERT pre-trained on scientific texts, demonstrated superior performance in named entity recognition (NER) and relationship classification (REL) tasks compared to its general-domain counterparts. By training on 1.14 million papers from the Semantic Scholar corpus, SciBERT outperformed BERT-BASE in both NER and REL tasks within the biomedical and computer science domains. This success highlights the importance of domain-specific training, as it enables models to understand specialized terms and concepts critical for correlating material names with properties.

Similarly, MatBERT[77] was developed to enhance the extraction of material-property pairs by fine-tuning on both materials science texts and general-purpose natural language inference (NLI) corpora. The model's tokenizer and pre-training on materials science texts allowed it to grasp nuanced terms like "chalcogenide" and "band gap," essential for understanding the relationships between materials and their properties. This approach eliminated the need for labor-intensive feature engineering, enabling end-to-end extraction pipelines that are both efficient and effective.

A significant breakthrough in polymer informatics came with the development of PolyBERT[78], a chemical language model specifically designed for this field. PolyBERT streamlines the discovery of application-specific polymers by replacing traditional handcrafted fingerprinting methods with an end-to-end machine-driven pipeline. The model was pre-trained on 100 million hypothetical polymers, generated through combinatorial reassembly of 4,424 chemical fragments derived from 13,766 known polymers using the BRICS decomposition method. For fine-tuning, the authors utilized 13,766 experimentally characterized polymers with properties such as glass transition temperature (Tg), Young's modulus, and tensile strength. Built on the DeBERTa architecture, PolyBERT employs 12 Transformer encoders with 12 attention heads each, converting Polymer Simplified Molecular-Input Line-Entry System (PSMILES) strings into embeddings via token-level averaging from the final Transformer layer. This architecture enables PolyBERT to generate polymer fingerprints in just 10 milliseconds per polymer—100 times faster than traditional handcrafted methods—while maintaining accuracy, with mean absolute errors (MAE) of 21.4 K for Tg and 0.496 GPa for E, comparable to state-of-the-art approaches. By treating PSMILES as a "chemical language," PolyBERT eliminates manual feature engineering and facilitates scalable cloud deployment, significantly accelerating high-throughput screening.

ChemBERTa[79] is another compelling transformer-based model designed for molecular property prediction, inspired by the success of transformer architectures in natural language processing. Built on the RoBERTa framework, ChemBERTa utilizes a masked language modeling approach to learn molecular representations from a vast dataset of 77 million SMILES strings sourced from PubChem. This pretraining enables the model to capture complex chemical structures effectively. Fine-tuned on various MoleculeNet classification tasks, ChemBERTa demonstrated competitive performance, particularly as the size of the pretraining dataset increased, highlighting the model's ability to generalize across different tasks. Although it did not surpass the performance of established GNN models, ChemBERTa's scalable and efficient processing of large datasets shows significant promise for future applications in cheminformatics, offering a robust framework for leveraging unlabeled chemical data to enhance predictive accuracy.

The development and deployment of these domain-specific language models underscore the potential of LLMs to automate materials discovery. They offer a robust framework for rapid polymer property prediction and design, drastically reducing the time and resources needed for experimental validation. This capability is particularly advantageous in the fast-paced field of energy materials, where the demand for novel polymers with specific properties is high. Moreover, these models can be integrated with existing databases and experimental data to refine predictions and guide experimental designs. The ability of LLMs to continuously learn from new data ensures that they remain relevant and accurate as new materials and technologies emerge. This adaptability is crucial for staying ahead in the rapidly evolving field of polymer research.

In conclusion, LLMs have emerged as a cornerstone of modern literature mining, offering unparalleled scalability and contextual awareness in extracting synthesis-property relationships. By combining domain-specific fine-tuning, multimodal integration, and human-AI collaboration, these models are bridging the gap between unstructured scientific knowledge and structured materials databases. However, ongoing challenges—including hallucination mitigation, domain adaptation, and explainability—require concerted efforts from the materials informatics community. As LLMs evolve, their synergy with experimental and computational workflows promises to accelerate the discovery of next-generation polymers, enabling breakthroughs in energy storage, biomedicine, and sustainable materials.

### 3.3 High-Throughput Experimental Platforms

In the realm of AI-driven polymer research, experimental data serves as a critical pillar, bridging theoretical predictions and real-world material performance. The rise of inverse design strategies in polymer science has intensified the demand for experimental frameworks capable of testing computationally proposed materials at scale. Traditional experimental approaches, while foundational, often produce sparse or fragmented datasets due to manual execution, limited parameter exploration, and slow iteration cycles. For instance, synthesizing a single shape-memory polymer (SMP) with subsequent property measurements can span months, rendering the assembly of modest datasets (~100 data points) a multi-year endeavor – a timescale misaligned with modern materials discovery needs. Unlike inorganic materials, which benefit from massive theoretical libraries, polymer science faces unique challenges due to the stochastic nature of macromolecules. High-throughput experimentation (HTE) addresses this gap by enabling parallelized synthesis, purification, and characterization of polymers, systematically generating large-scale, standardized datasets to bridge computational proposals with empirical validation at unprecedented scales.[80,81]

Three interconnected strategies dominate contemporary HTE platforms, each addressing distinct aspects of polymer complexity. The 96-well-plate system, adapted from pharmaceutical screening, allows parallel synthesis of hundreds of polymer samples in microliter volumes. This method excels in combinatorial studies—such as optimizing monomer ratios or catalyst concentrations—while generating spatially organized datasets ideal for ML training. Recent advances demonstrate its compatibility with controlled electrodeposition of polymer films. The PANDA[82] system employs a novel 96-well plate architecture to accelerate polymer research through parallelized electrodeposition and multimodal characterization. Each well integrates a transparent indium tin oxide (ITO)-coated glass bottom, serving as a working electrode for electrochemical deposition while enabling real-time optical transmission measurements. This setup allows simultaneous testing of 96 unique polymer formulations, with automated dispensing and electrochemical control. However, limitations in heat dissipation and post-synthesis purification restrict its utility for exothermic reactions or multi-step polymerizations.

Flow chemistry addresses these thermal challenges through continuous reagent mixing in tubular reactors, enabling precise temperature control and inline spectroscopic characterization.[83] Yet, viscosity changes during chain growth often confine flow systems to low-molecular-weight polymers, limiting their scope for high-performance polymer discovery.[83,84] Recently, Zhou[85] et al. develops a computer-controlled droplet-flow platform for photo-regulated polymerization, enabling rapid synthesis of copolymer libraries under high-viscosity conditions. By encapsulating polar monomer solutions (e.g., 78 wt% DMA in DMSO) into discrete droplets using a nonpolar carrier solvent (n-hexane), the system overcomes flow limitations caused by viscosity increases during polymerization. Automated software regulates flow rates and reaction parameters, producing 275 polymer samples in 11 minutes with consistent control. The method facilitates scalable, oxygen-tolerant synthesis of tailored (co)polymers and accelerates structure–property screening, exemplifying flow chemistry's potential for high-throughput material discovery without batch-mode constraints.

Robotic platforms represent the pinnacle of HTE synthesis, combining precision and scalability. For instance, this system can synthesize hundreds of polymer variants in a single campaign—precisely tuning variables like monomer ratios, catalysts, and reaction conditions—while integrated analytical tools such as high-throughput Fourier-transform infrared spectroscopy (FTIR) or gel permeation chromatography (GPC) measure properties like molecular weight distribution, thermal stability, or rheological behavior.[86] This approach not only accelerates data generation but also ensures consistency, minimizing human error and enabling direct comparability across samples. HTE's structured experimental designs, such as factorial or response surface methodologies, further enhance data utility for AI by explicitly mapping input parameters (e.g., temperature gradients, stoichiometric imbalances) to output properties, creating labeled datasets ideal for supervised learning.[87]

Michael[88] et al. introduces a robotic platform for rapid formulation and characterization of solid polymer electrolytes, accelerating research by 100x. The system automates dispensing, mixing, and casting of poly(ethylene oxide)-based films, followed by in-situ thickness measurement and variable-temperature electrochemical impedance spectroscopy. Testing 70+ formulations (330 samples, 2000+ data points), it reveals a universal colligative relationship between glass transition temperature minima and ionic conductivity maxima, independent of cation/anion chemistry. The platform processes 90 samples in 5 days, integrating automated DSC analysis for thermal properties. This approach enables large-scale, reproducible datasets critical for advancing sodium- and lithium-ion battery electrolytes.

The integration of HTE with autonomous experimental platforms represents a paradigm shift toward "self-driving labs". These platforms combine robotics, real-time analytics, and AI-driven decision-making to form closed-loop workflows where experiments are designed, executed, and analyzed iteratively without human intervention. For example, Bayesian optimization algorithms can navigate multi-dimensional parameter spaces—such as monomer combinations, solvent polarities, and curing times—prioritizing experiments that maximize information gain or target predefined material performance thresholds.[89–93] Real-time feedback from inline sensors (e.g., microfluidic viscometers) or automated characterization tools enables dynamic adjustments, allowing the system to refine hypotheses on-the-fly.

Notable implementations include platforms like Polybot[94], which has demonstrated the synthesis and optimization of hundreds of electronic polymers within weeks, a task that would span years using conventional methods. These systems also synergize with computational databases, cross-referencing experimental results with molecular dynamics simulations or literature data to validate predictions or resolve inconsistencies, thereby enhancing both experimental efficiency and computational accuracy. Jurğis[6] et al. demonstrate an autonomous platform for polymer electrolyte discovery by integrating warm-start Bayesian optimization (BO) with robotic experimentation. Their platform leverages a chemistry-informed neural network, ChemArr, pretrained on literature data encompassing 229 polymers and 41 lithium salts, to guide the Bayesian optimization. The study systematically explored 51 unique PCL-salt combinations across 17 lithium salts—a 4-fold expansion in salt diversity compared to prior literature. Each formulation underwent rigorous triplicate testing at four temperatures (30–90°C), generating 710 standardized data points that expanded existing PCL electrolyte databases by 10×. This high-throughput approach not only accelerated discovery but ensured precision through robotic synthesis, inline characterization (e.g., impedance spectroscopy), and blockchain-secured metadata annotation.

While HTE and autonomous systems offer unprecedented data scalability, challenges persist in data harmonization and quality control. HTE generates heterogeneous datasets from diverse instruments—e.g., spectral data from nuclear magnetic resonance (NMR), stress-strain curves from tensile testers, and thermal degradation profiles from differential scanning calorimetry (DSC)—which require robust preprocessing pipelines to standardize formats and eliminate noise. Metadata annotation, including experimental conditions, instrument calibration details, and environmental factors, must be rigorously standardized to ensure reproducibility and model interpretability. Emerging solutions such as blockchain-secured digital lab notebooks and domain-specific ontologies like Polymer Ontology (PO) are addressing these gaps by unifying terminology and enabling traceability. Furthermore, the fusion of HTE with autonomous platforms demands scalable computational infrastructure, including cloud-based data storage, GPU-accelerated analysis, and low-latency communication between robotic hardware and AI controllers.

Looking ahead, the convergence of HTE, autonomy, and AI promises to redefine polymer discovery. Future "cloud labs" could enable remote AI systems to orchestrate geographically distributed HTE platforms, facilitating 24/7 experimentation and global collaboration. Federated learning frameworks may allow decentralized training of AI models on proprietary experimental datasets while preserving data privacy—a critical feature for industrial R&D. Additionally, the integration of generative AI with HTE could enable de novo polymer design: models like GNoME (Google's Graph Networks for Materials Exploration) might propose novel monomers or architectures, which autonomous platforms then synthesize and validate iteratively. These advancements will democratize access to high-quality experimental data, empowering smaller academic labs to compete with industrial-scale facilities in the AI-driven polymer innovation landscape. By merging the scalability of HTE with the adaptive intelligence of autonomous systems, researchers are poised to unlock previously inaccessible regions of chemical space while aligning with sustainability goals through resource-efficient experimentation.

### 3.4 Data Generation by Computational Modeling and Machine Learning

The integration of computational modeling and machine learning (ML) into polymer research has emerged as a transformative paradigm for generating high-fidelity datasets, particularly in domains where experimental data remain sparse or resource-intensive to acquire. Unlike inorganic materials, polymers pose unique challenges for computational analysis due to their structural complexity, which spans amorphous, semicrystalline, or hybrid phases. These complexities necessitate multi-scale approaches that balance accuracy with computational feasibility. While traditional experimental methods struggle with the synthetic difficulty and time-consuming characterization of polymers, computational tools—ranging from quantum mechanical simulations to mesoscale modeling—offer a systematic pathway to predict and optimize material properties. However, the reliability and scalability of these methods depend on the synergy between physics-based models and data-driven algorithms, which together enable the exploration of vast chemical and structural landscapes.

One of the most prominent computational frameworks in polymer science is density functional theory (DFT), which provides atomic-scale insights into electronic and dielectric properties. For instance, DFT calculations have been instrumental in estimating charge injection barriers, bandgaps, and trap depths— parameters critical for predicting breakdown strength and conduction loss in dielectric polymers.[95–97] These properties are notoriously difficult to measure experimentally due to the sensitivity of polymers to defects and interfacial interactions. By modeling small-scale systems (<100 atoms), DFT bypasses these experimental bottlenecks, enabling researchers to correlate molecular structures with macroscopic behaviors. For example, studies have demonstrated how variations in polymer side chains or backbone rigidity influence charge transport mechanisms, offering design rules for high-performance dielectrics.[95,98] However, the computational expense of DFT restricts its application to simplified models, such as single-chain or idealized crystalline structures, which may not fully capture the heterogeneity of real-world polymers. To address this limitation, hierarchical modeling strategies have been developed. The polymer structure predictor (PSP) workflow, for instance, generates single-chain, crystalline, and amorphous models to approximate realistic polymer morphologies. This approach has enabled the creation of the Khazana database, a rare computational resource containing DFT-derived properties like bandgaps, dielectric constants, and refractive indices for thousands of polymers. Such databases not only supplement experimental datasets but also serve as training grounds for ML models aimed at accelerating material discovery.

Molecular dynamics (MD) simulations complement DFT by bridging atomic-scale interactions with mesoscale phenomena. Classical MD, parameterized with empirical force fields, is widely used to study thermodynamic and mechanical properties, including glass transition temperatures (Tg), thermal conductivity, and tensile strength.[99–102] For example, non-equilibrium MD simulations have been employed to compute thermal conductivity in polymer nanocomposites, revealing how nanofiller orientation and dispersion influence heat transfer.[103,104] Despite these successes, MD faces challenges in simulating large systems (>10,000 atoms) over experimentally relevant timescales. For instance, simulating relaxation effects in shape-memory polymers (SMPs) requires nanosecond-scale trajectories, which often underestimate experimental measurements by a factor of two due to truncated simulation times.[105,106] Moreover, the accuracy of MD hinges on the availability of reliable force fields, which are scarce for novel polymer chemistries. Recent advances in machine-learned force fields promise to mitigate this issue by training on ab initio data, thereby extending MD's predictive power to unconventional polymers.

At the mesoscale, phase-field modeling and finite-element methods (FEM) enable the study of polymer nanocomposites and bulk material behaviors. Phase-field models excel in capturing microstructure-dependent properties, such as effective permittivity and breakdown pathways in dielectric nanocomposites. By simulating nanofiller shapes, orientations, and interfacial interactions, these models provide design principles for enhancing energy storage density.[107,108] FEM, on the other hand, is indispensable for analyzing large-scale phenomena like space charge distribution and thermal gradients, which are critical for applications in capacitors and insulation materials. For example, bipolar charge transport models implemented via FEM have elucidated how space charge accumulation near electrodes precipitates dielectric failure—a phenomenon that experimental techniques struggle to observe in real time.[109,110] These multi-scale approaches are increasingly integrated with ML to identify optimal microstructures or processing conditions. Shen et al., for instance, combined high-throughput phase-field simulations with ML to screen polymer nanocomposites for superior energy storage performance, demonstrating how computational workflows can prioritize candidate materials for experimental validation.[111]

The rise of autonomous computational platforms marks a significant leap toward data-driven polymer design. These systems leverage ML models trained on existing datasets to predict properties for hypothetical polymers, iteratively refining their predictions through active learning. For example, an autonomous agent developed by Huan[112] et al. dynamically selects polymers with target electronic properties, generates their 3D structural models using PSP, and validates predictions through DFT calculations. This closed-loop approach not only accelerates discovery but also enriches the computational dataset, creating a virtuous cycle of model improvement. Such platforms are particularly valuable for multi-objective optimization, where balancing competing properties—such as high thermal conductivity and low dielectric loss—requires exploring vast chemical spaces.

Despite these advancements, computational methods face intrinsic limitations. First-principles calculations, while accurate, are prohibitively expensive for large systems or high-throughput screening. Semiempirical models and simplified theoretical equations. Moreover, the interpretability of ML models remains a challenge; black-box predictions hinder the extraction of actionable design rules. Future efforts must focus on hybrid frameworks that integrate physics-based constraints into ML algorithms, ensuring predictions align with fundamental principles. Additionally, community-wide initiatives to standardize computational protocols and share datasets—akin to the Materials Project for inorganic materials—are essential to overcome fragmentation in the polymer field.

In summary, computational modeling and ML are indispensable tools for generating polymer data at scale, particularly for properties that are experimentally elusive. By combining quantum mechanical accuracy, mesoscale dynamics, and data-driven optimization, these methods unlock new avenues for material discovery. However, their full potential will only be realized through continued innovation in multi-scale modeling, force field development, and collaborative data sharing—a vision that aligns with the broader shift toward open science and accelerated materials development.

## 3.5 Crowdsourcing and Community-driven Database Creation

The integration of crowdsourcing into polymer data collection represents a transformative shift that complements and enhances traditional methods such as database curation, computational modeling, literature mining, and experimental campaigns. While these conventional approaches provide foundational datasets, they often face limitations in capturing nuanced synthesis protocols, failure modes, or contextual metadata that are critical for energy material optimization. Crowdsourcing bridges these gaps by leveraging collective intelligence from academia and industry, fostering a dynamic ecosystem where researchers collaboratively contribute, validate, and refine data. This paradigm not only expands the scope and diversity of datasets but also improves accuracy through community-driven validation, addressing ambiguities inherent in automated text mining or fragmented experimental records.

Platforms like the Polymer Property Predictor and Database (PPP&D)[60,113] exemplify this synergy, aggregating experimentally validated data on ionic conductivity, thermal stability, and mechanical properties through researcher contributions. Unlike purely computational or literature-derived datasets, such platforms incorporate contextual details—such as solvent annealing time or plasticizer content—that are frequently omitted in publications but crucial for reproducibility. Similarly, NanoMine[114] employs hybrid methodologies, combining data extraction from literature with crowdsourced annotations to resolve ambiguities in polymer nanocomposite data (e.g., distinguishing homopolymer blends from copolymers). These initiatives highlight how crowdsourcing complements text mining by enriching structured data with tacit knowledge, such as lab-specific synthesis quirks or instrument calibration details, which automated systems often overlook.

Central to these efforts is the development of adaptive schema frameworks that balance standardization with flexibility. Drawing lessons from biology's Protein Data Bank (PDB)[115], which standardizes 38,000+ protein structures through rigorous schemas, polymer informatics faces the added complexity of diverse material classes (e.g., solid electrolytes, ionomers) requiring tailored metadata. Emerging solutions like the Materials Data Curation System (MDCS)[116] enable researchers to select or co-create schemas from a centralized library, accommodating everything from copolymer sequence distributions to operando battery cycling conditions. This modular approach mitigates the "schema bottleneck" by allowing contributors to define context-specific parameters (e.g., "plasticizer content" or "solvent annealing time") while maintaining interoperability through shared ontologies. The Materials Data Facility[117] (MDF) further accelerates this process by hosting datasets of all scales—from individual lab measurements to multi-institution campaign data—with machine-readable descriptors that link directly to AI training pipelines.

Incentivizing data sharing remains pivotal to scaling these ecosystems. While journal mandates (e.g., Science's structure data requirements) and funding agency policies (NSF's data management plans) provide foundational pressure, sustainable engagement requires intrinsic motivation. Platforms are increasingly adopting contributor credit systems modeled after open-source software communities, where dataset citations count toward academic metrics, and interactive tools like the Galaxy Project lower entry barriers by integrating data submission with analysis workflows. Early successes include the PolyDAT[118] initiative, which rewards contributors with prioritized access to AI-driven property prediction models trained on their shared data—a virtuous cycle that enhances both dataset richness and model accuracy.

Challenges persist in harmonizing heterogeneous data formats and ensuring quality control. Advanced validation layers now employ AI-assisted outlier detection, where machine learning models flag outliers in the database based on established structure-property relationships.[119–122] Blockchain-inspired timestamping mechanisms enhance data provenance[123], while repositories like the Schema Registry[124] archive version-controlled schemas to ensure reproducibility amid evolving standards. Looking ahead, the integration of crowdsourced databases with autonomous experimentation systems—where robotic labs directly upload synthesis outcomes to shared repositories—promises to close the loop between data generation and application, enabling real-time updates and iterative design.

By transcending the limitations of isolated data collection methods, crowdsourcing emerges not merely as a supplementary tool but as a cornerstone of next-generation polymer informatics. It unites human expertise with machine intelligence, creating a collaborative framework that accelerates the discovery of sustainable energy materials while ensuring datasets are both comprehensive and contextually robust. This approach positions the scientific community to tackle complex challenges—from solid-state battery electrolytes to recyclable thermosets—with unprecedented agility and precision.

## 4. Future Directions and Outlook

The rapid evolution of polymer science, driven by the urgent demand for advanced energy materials, necessitates a paradigm shift in how data ecosystems are designed, governed, and utilized. While significant progress has been made in addressing data fragmentation, quality, and standardization challenges, the road ahead requires innovative strategies that integrate emerging technologies, interdisciplinary collaboration, and systemic reforms. This section outlines critical future directions to bridge existing gaps and unlock the full potential of polymer informatics as a cornerstone of energy material innovation.

**Interoperability Through AI-Driven Ontologies and Semantic Frameworks**: A foundational step toward resolving data silos lies in the development of universal, machine-actionable ontologies tailored to polymer science. Current efforts, such as the Polymer Genome Initiative and FAIR (Findable, Accessible, Interoperable, Reusable) data principles, provide a starting point but lack the granularity required for energy-specific applications. Future ontologies must incorporate domain-specific attributes—such as ion transport mechanisms in solid-state electrolytes or degradation pathways in photovoltaic polymers—while aligning with broader materials science frameworks. Advances in natural language processing (NLP) and large language models (LLMs) like GPT-4 and polyBERT could automate the extraction of context-aware metadata from legacy literature, experimental protocols, and computational outputs, enabling dynamic ontology updates. For instance, LLMs trained on polymer synthesis papers could infer implicit relationships between processing conditions and mechanical properties, enriching structured databases with previously unstructured knowledge. Semantic graph databases, which map complex relationships between monomers, synthesis methods, and performance metrics, will likely replace traditional relational databases, enabling federated queries across academic, industrial, and governmental repositories.

**Democratizing Data Access via Decentralized Infrastructure**: The centralized data repositories dominating polymer science today—often constrained by proprietary barriers or institutional boundaries—will gradually give way to decentralized architectures leveraging blockchain and federated learning systems. Blockchain-based platforms could ensure data provenance and incentivize sharing through smart contracts, where contributors receive cryptographic tokens for uploading high-quality datasets or validating metadata. Federated learning, which trains machine learning models across distributed datasets without transferring raw data, would mitigate privacy concerns in industrial-academic collaborations. For example, a pharmaceutical company could collaboratively improve battery binder models using proprietary electrolyte data while keeping sensitive information on-premises. Such systems must be paired with lightweight, open-source tools for data annotation and quality control to empower researchers in resource-limited settings. Emerging initiatives like the Materials Data Facility and the Battery Data Genome Project exemplify early steps in this direction, though scaling these to global polymer communities remains a formidable challenge.

**Closing the Loop: Autonomous Laboratories and Self-Optimizing Systems**: The integration of high-throughput experimentation (HTE), machine learning, and robotic automation will evolve into fully autonomous laboratories capable of self-directed discovery. These systems, already prototyped in organic photovoltaics and solid-state battery research, will generate self-consistent datasets by iteratively designing experiments, synthesizing materials, and characterizing properties with minimal human intervention. For instance, a robotic platform guided by Bayesian optimization could explore copolymer compositions for fuel cell membranes, automatically recording synthesis parameters (e.g., reaction temperature, catalyst loading) and correlating them with ion conductivity measurements. Crucially, such systems must embed data quality checks at every stage—using computer vision to detect synthesis anomalies or ML models to flag outliers in spectroscopic data. Over time, autonomous labs could feed data into physics-informed neural networks (PINNs), which combine empirical results with fundamental equations (e.g., Fick's law for diffusion in polymer electrolytes) to extrapolate beyond experimentally accessible conditions. This closed-loop paradigm would not only accelerate discovery but also systematically reduce biases inherent in human-driven experimental design.

**Ethical AI and Equitable Data Governance**: As AI becomes deeply embedded in polymer informatics, ethical considerations must transition from afterthoughts to core design principles. Current AI/ML models often perpetuate biases—for example, overrepresenting commercially popular polymers like polyethylene terephthalate (PET) in training data, thereby limiting their predictive power for novel biopolymers. Future frameworks should mandate diversity audits for training datasets and incorporate uncertainty quantification to flag predictions extrapolated beyond reliable data ranges. Equally critical is addressing the "data divide" between well-resourced institutions and smaller research groups. Open-access platforms funded by international consortia, akin to the Human Genome Project, could pool resources for large-scale data generation while ensuring equitable access. Licensing models like Creative Commons with reciprocity clauses—where commercial users contribute derived data back to the public domain—might balance innovation incentives with collective progress. Regulatory bodies will need to establish standards for AI accountability in materials discovery, particularly for safety-critical applications like nuclear waste encapsulation polymers or implantable bio-batteries.

**From Human-Centric to Hybrid Intelligence Systems**: The future of polymer data ecosystems lies in hybrid intelligence systems that synergize human expertise with machine capabilities. While LLMs excel at extracting patterns from vast literature corpora, they lack the nuanced domain knowledge of seasoned polymer chemists. Hybrid systems could embed human feedback loops—for example, allowing researchers to correct misannotated Tg values in automated literature mining outputs or prioritize understudied polymer classes for HTE campaigns. Citizen science initiatives, gamified through platforms like Foldit (adapted for polymer folding challenges), might engage broader communities in data curation tasks while educating the public on energy materials. Additionally, virtual reality (VR) interfaces could enable immersive exploration of polymer structure-property landscapes, helping researchers intuitively grasp complex relationships obscured by traditional 2D visualizations.

**Preparing for Exascale Data Challenges**: The advent of quantum computing and exascale classical computing will reshape polymer data generation and analysis. Quantum simulations, capable of modeling electron transport in conjugated polymers with sub-angstrom precision, will generate datasets orders of magnitude larger than today's DFT (density functional theory) calculations. Managing these data deluges will require ultra-efficient compression algorithms and in-situ analysis tools that filter relevant features in real time. Meanwhile, quantum machine learning models could uncover non-linear relationships in multi-dimensional datasets—for instance, linking subtle variations in monomer sequence to ionic conductivity in block copolymer electrolytes. However, realizing this potential demands closer collaboration between polymer scientists and quantum algorithm developers to translate domain-specific problems into quantum-native representations.

The data challenges facing polymer science for energy applications are neither insurmountable nor unique—they reflect broader growing pains in the transition to data-driven discovery. Success hinges on building adaptive, ethically grounded ecosystems that leverage AI not as a replacement for human ingenuity but as a catalyst for collective intelligence. By prioritizing interoperability, decentralization, and equity, the polymer community can transform its data landscape from a patchwork of isolated efforts into a globally connected knowledge infrastructure. The ultimate goal is clear: a future where every synthesis protocol, computational simulation, or failed experiment contributes to an ever-refining tapestry of understanding, accelerating the development of polymers that power sustainable energy systems. This vision demands bold investments in technology, policy, and education today to ensure that the polymer data revolution leaves no molecule—and no researcher—behind.

**Aknowledgement:**

We gratefully acknowledge the financial support from Hong Kong Quantum AI Lab Limited, Air @ InnoHK of Hong Kong Government.